\documentstyle[aps,prb,graphicx,rotating]{revtex}

\begin{document}

\draft

\title{Fermionic Heisenberg Model for Spin Glasses with BCS Pairing Interaction}

\author{S.\ G.\ Magalh\~{a}es\footnote{e-mail: ggarcia@ccne.ufsm.br} 
        and A.\ A.\ Schmidt\footnote{e-mail: alex@lana.ccne.ufsm.br}}

\address{Departamento de Matem\'{a}tica, CCNE\\
         Universidade Federal de Santa Maria\\
         97105-900 Santa Maria, RS, Brazil\\}
	 
\date{\today}

\maketitle

\begin{abstract}
In the present paper we have analysed a fermionic infinite-ranged quantum 
Heisenberg spin glass ($s=1/2$) with a BCS coupling in real space in the
presence of an applied magnetic field. This model has been obtained by 
tracing out the conducting fermions in a superconducting alloy. The magnetic 
field is applied in the resulting effective model. The problem is formulated 
in the path integral formalism where the spin variables are defined as bilinear
combinations of the Grassmann fields. The static approximation is used to treat
both the pairing and the spin terms together with the replica symmetry ansatz. 
Henceforth, the problem can be reduced to a one site problem. The field in the 
$z$ direction, $H_{z}$, separates the order parameters in two groups: parallel 
and transverse to it. We have obtained a phase diagram in $T$--$g$ space with 
zero transverse spin-glass ordering, $g$ being the strength of the pairing 
interaction. It has been possible to locate the transition temperature between 
the normal paramagnetic phase (NP) and the phase where there is a long range 
order corresponding to formation of pairs (PAIR). The transition ends at the 
temperature $T_{f}$, the transition temperature between the NP phase and the 
spin glass (SG) phase. $T_{f}$ decreases for stronger fields allowing us to 
calculate the NP--PAIR line transition even at low temperatures. The NP--PAIR 
transition line has a complex dependence with $g$ and $H_{z}$, having a 
tricritical point depending on $H_{z}$ from where second order transitions 
occur for higher values of $g$ and first order transitions occur for lower 
values of $g$. 
\end{abstract}



\section{Introduction} \label{intro}

The interplay between a magnetic order and superconductivity is a current 
issue. Particularly, spin glass ordering has been reported in many physical 
systems that includes high-$T_{c}$ superconductors\cite{1,2}, heavy 
fermions\cite{3,4} and conventional superconductors doped with magnetic 
impurities\cite{5} like $Gd_{x}Th_{1-x}RU_{2}$. These conventional 
superconductors can be modeled by a s-d exchange interaction between the 
localized magnetic impurities and the conducting fermions which have 
conventional BCS interaction (Nass et al.\cite{6}). The main interest of 
Nass et al.\ was to find the density of states in the presence of magnetic 
impurities. However, relatively little consideration has been given to the 
phase transition problem between spin glass and a superconductive 
order\cite{She,Jose,Opp}.

Nevertheless, from a wider perspective the role of quantum fluctuations in 
disordered strongly interacting systems still needs to be fully understood. 
Theoretical studies in recent years have shown plenty of interesting results. 
For instance, the presence of a non-Fermi liquid behavior near $T=0$ 
transition between a metallic paramagnetic and a metallic spin glass in a 
model introduced to study the competition between Kondo effect and RKKY 
interaction\cite{Ge}. In addition, the random quantum Heisenberg with 
generalized SU(M) spins\cite{Sach} produces a spin liquid ground state in 
the large M limit with a local dynamic susceptibility that is identical to 
a marginal Fermi Liquid. 

Recently, a model has been introduced to study the phase transition between 
the spin glass ordering and the BCS pairing among fermions of opposite 
spins\cite{7}. This model has been obtained from the model of Nass et 
al.\cite{6} by tracing out the conducting fermionic degrees of freedom by 
perturbation expansion. The remaining effective problem could be solved in 
its infinite ranged Ising version. A rich phase diagram has been found in 
the $T$--$g$ plane ($g$ is the strength of the pairing interaction) for 
fixed $J$ (variance of the spin random coupling $J_{ij}$) with a normal 
paramagnetic phase (NP), a spin glass phase (SG) and a pairing phase (PAIR) 
where there is formation of pairs in the sites. The first two phases (NP 
and SG) are separated by a second order line at the spin glass transition 
temperature $T_{f}$. However, the line $g=g_{c}(T)$ between the NP and PAIR 
phases is more complex. It is a second order line which ends at a tricritical 
point ($T_{tc}, g_{tc}$), where a first order line starts. Below the spin 
glass transition temperature $T_{f}$, the last two phases (SG and PAIR) are 
separated by the same first order line transition.   

The purpose of the present study is to investigate the infinite ranged 
Heisenberg version of the model given by Ref.\ \onlinecite{7} 
in presence of a magnetic field for the half filling case. This many fermion 
problem is formulated by defining the SU(2) spin variables as a bilinear 
combination of Grassmann fields and computing the partition function 
through the path integral formalism. The disorder in the coupling $J_{ij}$ 
is treated by standard replica formalism. For this representation of the 
spin variable, the Fock space ($s=1/2$) has four quantum states per site 
but only two of them are magnetic. As a consequence, the sensitivity of 
the site to magnetic coupling is dependent on the occupation number which 
is controlled by quantum statistics. When the strength of the BCS coupling 
is increased, forcing the double occupation, the number of sites 
non-sensitive to the magnetic coupling increases. Therefore, this 
mechanism could suppress magnetic ordering such as the spin glass phase. 
Actually, that is what happens in the Ising model\cite{7} even at the half 
filling situation. 
  
In the replica formalism for the quantum spin glass, the replica diagonal 
order parameter has to be solved coupled with the non-diagonal one (the 
spin glass order parameter) and the remaining order parameters of the 
problem. In consequence, the diagonal component is no longer constrained 
to unity and has an important role in the problem. In fact, the location 
of $T_{f}$ can be obtained through this component. This is the essential
point: the location of other transitions of the problem are also dependent 
of this diagonal component even when the non-diagonal component is zero.  

Another important aspect of these problems is that the order parameter
(the spin-spin correlation function) are time dependent. Bray and More\cite{8}, 
in their pioneering work, investigated the role of the quantum 
fluctuations in the Heisenberg spin glass model by using the static 
approximation. They formulated the problem in the Feynman's path integral 
formalism where the spins variables have a dependence on fictitious 
time $\tau$. The static approximation neglects the time dependence giving an 
upper bound for the thermodynamic potential and the location of the 
transition points. For temperatures above the $T_{f}$, this approximation is 
expected to be exact together with the replica symmetric solution. This same 
problem has been studied\cite{Alb} with a formalism where the spin variables 
were represented by bilinear combinations of Grassmann fields. The static 
approximation was used for the spin glass order parameter, while for the 
susceptibility (associated to the replica diagonal order parameter) was 
applied an instantaneous approximation. This approach has found a non-zero 
spin glass transition temperature for $s=1/2$. Goldschmidt and Pik-Yin 
Lai\cite{9} used the same treatment as Bray and More\cite{8} to obtain a 
phase diagram for the infinite ranged quantum Heinsenberg and XY spin glass 
in an external magnetic field. Due to the magnetic field, the order 
parameters were separated in two groups: transversal to the field and 
parallel to it. This approach allowed to locate the spin glass transition 
transversal to the field and investigate the role of the tunneling in the 
transition. For instance, the results showed  that the spin glass ($s=1/2$) 
transition temperature $T_{f}$ is depressed for the Heisenberg model as the 
field strength is increased. Nevertheless, the tunneling was not strong 
enough to produce a transition at $T=0$.

In the present fermionic formulation, we have combined the methods of Refs.\
\onlinecite{Alb} and \onlinecite{9} by applying a field over the fermionic 
spin variables,  taking the transversal spin glass order parameter as null 
and using the static approximation for the remaining order parameters. In 
these circumstances, the magnetic field also has depressed the transversal 
spin glass transition temperature. This approach has allowed us to study the 
transition line between NP phase and the PAIR phase (we use here the same 
terminology of Ref.\ \onlinecite{7}) even in lower temperature. This 
transition line ends at the transversal spin glass transition temperature 
which moves down as the field strength enhances.

This paper is organized as follows. In section \ref{gen_form}, we present our 
model and develop it to obtain the Gran Canonical Potential and the saddle 
point equations for the order parameters in the half filling case. In section 
\ref{phase_diag}, we analyse the transition line between the NP and PAIR 
phases for several values of the magnetic field. The results show that as the 
magnetic field increases, larger values of the pairing interaction strength 
are necessary to produce a transition. We also locate the transition line 
between the normal paramagnetic phase and the spin glass phase. Discussions 
and concluding remarks are presented in the last section. 

   
\section{General Formulation} \label{gen_form}

We consider the following Hamiltonian\cite{7} in the presence of a 
magnetic field $H_{z}$:
\begin{eqnarray} 
\overline{H} = H-\mu N = -\sum_{ij}J_{ij}\left[\sigma^{z}_{i}\sigma^{z}_{j}
+\frac{1}{2}\left(\sigma^{+}_{i}\sigma^{-}_{j}+\sigma^{-}_{i}\sigma^{+}_{j}
\right)\right]   
-\mu\sum_{i}\sum_{s=\uparrow,\downarrow}c^{\dagger}_{is}c_{is}
-\frac{g}{N}\;\sum_{i,j}c^{\dagger}_{i\uparrow}c^{\dagger}_{i\downarrow}
c_{j\downarrow}c_{j\uparrow}-H_{z}\sum_{i}\sigma^{z}_{i}
\label{2.1}
\end{eqnarray}
where $\sigma^{\pm}_{i}=\sigma^{x}_{i}\pm \sigma^{y}_{i}$, 
the chemical potential is given by $\mu$ and $c^{\dagger}_{is}$ ($c_{is}$) 
are fermions creation (destruction) operators  ($s=\uparrow$ 
or $\downarrow$ means the spin projection). The spin operators 
are represented as bilinear combination of fermionic operators:
\begin{eqnarray} 
\sigma^{z}_{i}&=&\frac{1}{2}[c^{\dagger}_{i\uparrow}c_{i\uparrow}-
c^{\dagger}_{i\downarrow}c_{i\downarrow}] \\
\label{2.2}
\sigma^{+}_{i}&=&c^{\dagger}_{i\uparrow}c_{i\downarrow}\;\;.
\label{2.3}
\end{eqnarray}

The random coupling  $J_{ij}$ in Eq.\ (\ref {2.1}) are infinite 
ranged with a Gaussian distribution with mean and variance given by:
\begin{eqnarray} 
\langle J_{ij} \rangle_{ca}&=&0\nonumber\\
\langle J_{ij}^{2}\rangle_{ca}&=&4J^{2}/N
\label{2.4} 
\end{eqnarray}
where $N$ is the number of sites. 

The third term in Eq.\ (\ref{2.1}) represents a BCS pairing interaction
in real space. This pairing mechanism in fermionic problems already has been 
purposed, although, in different context\cite{10,11}. The remaining terms 
describe a fermionic Heisenberg spin glass in the presence of a field.
Our interest in this fermionic problem is to analyse the phase transitions at 
mean field level as firstly introduced by Sherrington and Kirkpatrick\cite{12} 
for classical spin glass, reducing the initial many site problem to a one site 
problem. In order to obtain that, the partition function is expressed in terms 
of functional integrals\cite{13} using the anti-commuting Grassmann variables 
$\phi^{\ast}_{is}(\tau)$ and $\phi_{is}(\tau)$ ($\tau$ is a complex time):   
\begin{equation} 
Z=\int D(\phi^{\ast}\phi)\,\exp[A_{0}+A_{pair}+A_{SG}+A_{field}]
\label{2.5}
\end{equation}
where the action $A_{0}$ in Eq.\ (\ref{2.5}) is the free part of the total 
action, $A_{pair}$ is the pairing part and $A_{SG}$ is the spin part.  

Using time Fourier transformed quantities, the free part is given by
\begin{equation} 
A_{0}=\sum_{i}\sum_{\omega} \left( i\omega +\beta \mu \right) 
\left[\phi _{i\uparrow}^{*}(\omega)\phi _{i\uparrow}(\omega)+
\phi _{i\downarrow}^{*}(\omega)\phi _{i\downarrow}(\omega) \right]
\label{2.6}
\end {equation}
and the $A_{field}$ is
\begin{equation} 
A_{field}=\frac{\beta H_{z}}{2}\sum_{i}\sum_{\omega}\left[\phi_{i\uparrow}^{*}
(\omega)\phi _{i\uparrow}(\omega)-\phi _{i\downarrow}^{*}(\omega)
\phi _{i\downarrow}(\omega)\right]
\label{2.7}
\end {equation}
while the pairing part of the action in the static approximation becomes
\begin{equation} 
A_{pair} \approx A_{pair}^{st}=\frac{\beta g}{N}\sum_{ij} 
\rho^{\ast}_{i}(0)\rho_{j}(0)
\label{2.8}
\end{equation}
where \(\displaystyle \beta=\frac{1}{T}\), $T$ is the temperature and
\begin{equation} 
\rho_{i}(0)=\sum_{\omega}\phi_{i\downarrow}(-\omega)
\phi_{i\uparrow}(\omega)
\label{2.9}
\end{equation}
with Matsubara's frequencies $\omega=(2m+1)\pi$, $m=0$, $\pm 1$, ...

Indeed, functional integral techniques within the static approximation have 
succeeded in studying phase transitions in conventional BCS 
superconductivity\cite{MUS} and in the presence of metal transition 
impurities\cite{ALB2}. It has been possible to build a mean field theory 
in momentum space by using a particle-hole transformation. Thus, the 
problem became exactly resolved. 

The spin part of the action is written as 
\begin{equation} 
A_H=\sum_{ij\nu }\frac{\beta J_{ij}}2\left\{S_i^z(\nu )S_j^z(-\nu)+
\frac{1}{2}\left[S_i^{+}(\nu)S_j^{-}(-\nu)+S_i^{-}(\nu)S_j^{+}
(-\nu)\right]\right\} 
\label{2.10}
\end{equation}
where the spins variables are expressed in terms of Grassmann fields:
\begin{eqnarray} 
S^{z}_{i}(\nu)&=&\frac{1}{2}\sum_{\omega}\left[\phi_{i\uparrow}^{*}(\omega+\nu)
\phi_{i\uparrow}(\omega)-\phi_{i\downarrow}^{*}(\omega+\nu)
\phi_{i\downarrow}(\omega)\right] \\
\label{2.11}
S^{+}_{i}(\nu)&=&\sum_{\omega}\phi^{*}_{i\uparrow}(\omega+\nu)
\phi_{i\downarrow}(\omega)\nonumber\\
S^{-}_{i}(\nu)&=&\sum_{\omega}\phi^{*}_{i\downarrow}(\omega+\nu)
\phi_{i\uparrow}(\omega)
\label{2.12} 
\end{eqnarray}
where $\nu=2\pi m$ in Eq. (\ref{2.12}).

~

The Grand Canonical Potential can be obtained through the replica method by
\begin{eqnarray} 
\Omega=-\frac{1}{\beta} \lim_{n \rightarrow 0} \frac{\langle Z^{n}
\rangle_{ca}-1}{n} .
\label{2.13}
\end{eqnarray}

The problem can be linearized and reduced to one site problem with the 
introduction of the auxiliary complex fields for the pairing part 
$\eta_{\alpha}(0)$, $\eta_{\alpha}^{*}(0)$ 
($\eta_{\alpha}(0)=\eta_{R \alpha}(0)+i\eta_{I \alpha}(0)$) 
and for the spin part\cite{Alb} 
$R_{\alpha\alpha}^{tt}(\nu,\nu^{\prime})=[R_{\alpha\alpha}^{tt}
(-\nu,-\nu^{\prime})]^{*}$ 
and 
$Q_{\alpha\beta}^{tt^{\prime}}(\nu,\nu^{\prime })
=[Q_{\alpha\beta}^{tt^{\prime}}(-\nu,-\nu^{\prime })]^{*}$ 
where $\alpha$ and $\beta$ are replica index and $t=x$, $y$, $z$. Thus
the configurational averaged replicated partition function is
\begin{eqnarray} 
\langle Z^n\rangle_{ca}&=&\frac{1}{{\aleph}^n}\int D(\eta^{*}\eta)\int 
D(Q^{*}Q)\int D(R^{*}R)\exp \left\{-\frac
N{8\beta^{2}J^{2}}\left[\sum_{\alpha}\sum_{t}{\mid}R_{\alpha\alpha}^{tt}
(\nu,\nu^{\prime}){\mid^{2}}+\right.\right.\nonumber\\
~&~&
\sum_{\alpha\beta}\sum_{t<t^{\prime}}
{\mid}Q_{\alpha\beta}^{tt^{\prime}}
(\nu,\nu^{\prime}){\mid^{2}}+ 
\sum_{\alpha <\beta}\sum_{t}{\mid}Q_{\alpha
\beta}^{tt}(\nu,\nu^{\prime}){\mid^{2}}+\left.\left.
\frac{N\beta g}4\sum_\alpha{\mid}\eta_\alpha(0){\mid^{2}}+
\ln(\Lambda)\right]\right\}
\label{2.14}
\end{eqnarray}
where \( \displaystyle \aleph= \left( \frac{2\pi}{N\beta^{2}J^{2}}\right) 
\left(\frac{\pi}{N\beta g}\right)\). 

~

The function $\Lambda$ in Eq.\ (\ref{2.14}) is
\begin{eqnarray} 
\Lambda&=&\int D(\phi ^{*}\phi )\exp\left\{A_o+\frac{\beta g}{2}
\sum_\alpha\left[\eta_\alpha(0)\rho ^{*}(0)+\eta_\alpha ^{*}(0)\rho 
(0)\right]+\beta H_z\sum_{i\alpha}S_{i\alpha }^z(0)\right.\nonumber\\
~&~&+\left[\sum_{i\alpha}\sum_{t}\sum_{\nu \nu ^{\prime }}
R_{\alpha\alpha}^{tt}(\nu ,\nu^{\prime })S_{i\alpha }^t(-\nu )
S_{i\alpha }^t(-\nu )\right.
+\sum_{i\alpha\beta}\sum_{t<t^{\prime }}\sum_{\nu \nu ^{\prime }}
Q_{\alpha\beta}^{tt^{\prime }}(\nu ,\nu^{\prime })S_{i\alpha }^t(-\nu 
)S_{i\alpha }^{t^{\prime }}(-\nu )\nonumber\\
~&~&+\left.\left.\sum_{i\alpha <\beta}\sum_{t}\sum_{\nu 
\nu ^{\prime }}Q_{\alpha \beta}^{tt}(\nu ,\nu ^{\prime })
S_{i\alpha }^t(-\nu )S_{i\beta}^{t}(-\nu )\right]\right\} .
\label{2.15}
\end{eqnarray}

Within the static approximation, together with the replica symmetric ansatz, 
the auxiliary fields of the spin glass part are giving by the following 
relations\cite{8}:
\begin{eqnarray} 
Q_{\alpha\beta}^{tt^{\prime}}(\nu,\nu ^{\prime}) &=&0 \;\;\;t\neq t^{\prime}
\nonumber\\
R_{\alpha\alpha}^{zz}(\nu ,\nu ^{\prime}) &=&4\beta ^2J^2R_z\delta _{\nu,\nu
^{\prime}}\delta _{\nu,0}\nonumber\\
R_{\alpha\alpha}^{xx}(\nu,\nu ^{\prime}) &=&R_{\alpha\alpha}^{yy}(\nu,
\nu^{\prime})=4\beta
^2J^2R\delta _{\nu,\nu ^{\prime}}\delta _{\nu ,0}\nonumber\\
Q_{\alpha \beta }^{xx}(\nu ,\nu ^{\prime }) &=&Q_{\alpha \beta }^{yy}(\nu,\nu
^{\prime})=4\beta^2J^2Q\delta_{\nu,\nu ^{\prime}}\delta_{\nu ,0}\nonumber\\
Q_{\alpha \beta }^{zz}(\nu,\nu ^{\prime}) &=&4\beta ^2J^2Q_z\delta_{\nu,\nu
^{\prime}}\delta_{\nu ,0} 
\label{2.16}
\end{eqnarray}
where $R$, $R_{z}$ and $Q_{z}$ are now real parameters. 

We also take the spin glass order parameter transversal to the field 
$Q=0$ to work in a temperature region where both static approximation and 
replica symmetric ansatz are reliable. The sum over the replica index gives 
again quadratic forms which are also linearized by new auxiliary fields. 
Thus, the one site $\Lambda$ function becomes
\begin{equation} 
\Lambda =
\int_{-\infty }^{+\infty}Dw\left\{\int_{-\infty}^{+\infty }Du_{x}
\int_{-\infty}^{+\infty }Du_{y}\int_{-\infty}^{+\infty }Dv
\int D(\phi ^{*}\phi )\exp [A_{0}\right.+
\left. A_{pair}+\beta\,\vec{h}\cdot\vec{S}]\right\}^n 
\label{2.17}
\end{equation}
where \( \displaystyle Dv =\frac{e^{-v^2/2}}{\sqrt{2\pi }}\,dv\) and
\begin{eqnarray} \label{2.18}
\vec{h}\cdot\vec{S}&=&h_{z}S_{z}+\frac{h_{+}S_{-}+h_{-}S_{+}}{2}\\
h_{\pm}&=&h_{x}\pm ih_{y}\;\; . \nonumber
\end{eqnarray}

The field $\vec{h}$ in the equation above is defined as:
\begin{eqnarray} \label{2.19}
h_{z}&=&J\left[v\sqrt{2(R_{z}-Q_{z})}+w\sqrt{2Q_{z}}+\frac{H_{z}}{2J}\right]\\
h_{+}&=&J\sqrt{2R}u_{+} \nonumber\\
h_{-}&=&J\sqrt{2R}u_{-}\;\; . \nonumber
\end{eqnarray}

The resulting problem obtained in Eq.\ (\ref{2.17}) is equivalent to fermions
in the presence of internal fields $g\eta/2$, $g\eta^{*}/2$ and $\vec{h}$. The 
two first ones are the pairing fields related to the long range order where 
there is pairing formation in the sites. The later is a random gaussian field 
related to the replica the components of the spin part of the auxiliary fields 
introduced in Eq.\ (\ref{2.14}), namely, the non-diagonal $Q_{z}$ and diagonal 
$R_{z}$ which are parallel to $H_{z}$, and diagonal $R$ which is transversal to 
$H_{z}$.

In order to construct a mean field theory which can be solved, we introduce 
the following matrices:
\begin{eqnarray}
\underline{\Psi}(\omega) = \left[ 
\begin{array}{c} 
\phi_{\uparrow}(\omega)        \\ \phi_{\downarrow}(\omega) \\
\phi_{\downarrow}^{*}(-\omega) \\ \phi_{\uparrow}^{*}(-\omega) 
\end{array}\right]
\label{2.20}
\end{eqnarray}
\begin{eqnarray} 
\underline{\Psi}^{\dagger}(\omega) = \left[ 
\begin{array}{cccc}
\phi_{\uparrow}^{*}(\omega) & \phi_{\downarrow}^{*}(\omega) & 
\phi_{\downarrow}(-\omega) & \phi_{\uparrow}(-\omega)
\end{array}
\right] 
\label{2.21}
\end{eqnarray}
and
\begin{eqnarray} 
\underline{G}^{-1}(\omega)=\left[ 
\begin{array}{cccc}
G_{\uparrow \uparrow }^{^{-1}}(\omega) & \beta h_{-} & \Delta & 0 \\ 
\beta h_{+} & G_{\downarrow \downarrow }^{^{-1}}(\omega) & 0 & -\Delta \\ 
\Delta ^{*} & 0 & -G_{\downarrow \downarrow }^{^{-1}}(-\omega) & 
-\beta h_{-} \\ 0 & -\Delta ^{*} & -\beta h_{+} & -G_{\uparrow 
\uparrow }^{^{-1}}(-\omega)
\end{array}
\right]\;\; .
\label{2.22}
\end{eqnarray}

In the equation above $G^{-1}_{ss}(\omega)=\sum_{s=\pm}(i\omega+\beta\mu+
s\beta h_{z})$ and \(\displaystyle \Delta =\frac{\beta g}{2}\eta\).  
Hence, for the $\Lambda$ function in Eq.\ (\ref{2.17}) we obtain
\begin{eqnarray} 
\Lambda =\int_{-\infty }^{+\infty }Dw\left\{\int_{-\infty}^{+\infty }Du_{x}
\int_{-\infty}^{+\infty }Du_{y}\int_{-\infty
}^{+\infty }Dv[I(w,u_{x},u_{y},v)]\right\}^n 
\label{2.23}
\end{eqnarray}
where
\begin{equation} 
I(w,u_{x},u_{y},v)=\int D(\phi ^{*}\phi )\exp\left[\frac{1}{2}
\sum_{\omega}\underline{\Psi}^{\dagger }(\omega)\underline{G}^{-1}(\omega)
\underline{\Psi}(\omega)\right]\;\; .
\label{2.24}
\end{equation}

The functional integral in Eq.\ (\ref{2.24}) can now be performed. The sum 
over Matsubara's frequencies in the resulting expression should be done as 
in Ref.\ \onlinecite{7}. The matrix formalism introduced in Eqs.\ (\ref{2.20}), 
(\ref{2.21}) and (\ref{2.22}) produces a particle-hole transformation in the 
fermions of spin down. Therefore, the Grand Canonical Potential can be found 
from Eqs.\ (\ref{2.10}) and (\ref{2.23}) as
\begin{equation} \label{2.25}
\Omega=4\beta J^2R^2+2\beta J^2R_z-2\beta J^2Q_z+\frac
g4|\eta| ^2 - \frac 1\beta \int_{-\infty }^{+\infty }Dw\ln(I_{\beta})
\end{equation}
where
\begin{equation}
I_{\beta} = \int_0^{+\infty}uDu\int_{-\infty}^{+\infty}Dv
\left[\cosh(\beta\mu^{\prime})+\cosh(\beta|\vec{h}|)\right]
\label{I_beta}
\end{equation}
and 
\begin{equation} 
\beta\mu^{\prime}=\sqrt{\beta\mu^{2}+\Delta^{2}}
\label{2.28}
\end{equation}
with $\vec{h}$ already defined in Eq.\ (\ref{2.19}). 

For \(\displaystyle S_{\beta}=\sinh(\beta|\vec{h}|)/(2\beta|\vec{h}|)\) and
$\theta=h_{z}/J$, the functions $R$, $Q_{z}$, $R_{z}$ and $|\eta|$ 
(from now on we write $\eta$ instead of $|\eta|$) can be determined from the 
saddle point equations that follow from Eq.\ (\ref{2.25}): 
\begin{eqnarray} \label{2.29}
R&=&\frac{1}{4}\int_{-\infty }^{+\infty}Dw\frac{\int_0^{+\infty
}u^3Du\int_{-\infty}^{+\infty}DvS_{\beta}} {I_{\beta}} 
\end{eqnarray}
\begin{eqnarray} \label{2.30}
R_z&=&\frac{1}{2\sqrt{(R_z-Q_z)}}\int_{-\infty}^{+\infty}Dw\frac{
\int_0^{+\infty}uDu\int_{-\infty}^{+\infty }vDvS_{\beta}\theta}{I_{\beta}}
\end{eqnarray}
\begin{eqnarray} \label{2.31}
Q_z&=&\frac{\beta^2\,J^2}{2}\int_{-\infty}^{+\infty}Dw
\left[\frac{\int_0^{+\infty}uDu\int_{-\infty
}^{+\infty}DvS_{\beta}\theta}{I_{\beta}}\right]^2 
\end{eqnarray}
\begin{eqnarray} \label{2.32}
\eta&=&\int_{-\infty}^{+\infty}Dw\frac{\sinh(\beta\mu^{\prime})}{I_{\beta}}\; .
\end{eqnarray}

We have solved numerically the set of Eqs.\ (\ref{2.29}), (\ref{2.30}),
(\ref{2.31}) and (\ref{2.32}) for the situation where there is one fermion 
per site (in average) which corresponds to $\mu=0$ due to the particle-hole 
transformation\cite{7}. 

The validity range in temperature for this theory is the region where
$T>T_{f}$ implying that it is necessary to locate the $T_{f}$ in this 
problem. This can be done by expanding the Grand Canonical Potential 
given in Eq.\ (\ref{2.14}) in powers of $Q^{tt}_{\alpha\beta}$ up to 
second order\cite{8,9}. The four spin correlation function that appears
in the coefficient of the quadratic term can be related to the parameter $R$ 
in the static approximation, where the averages are computed with an action 
where the auxiliary fields are given by Eq.\ (\ref{2.16}) and $Q=0$. 
Thus the temperature $T_{f}$ can be obtained from the condition that the
coefficient of the quadratic term vanishes, so we get  
\begin{equation} 
1=4\beta_{f}J R
\label{2.33}
\end{equation} 
along the set of Eqs.\ (\ref{2.29})--(\ref{2.32}).       


\section{Phase Diagram and Tricritical Point} \label{phase_diag}

The numerical solution of the equations (\ref{2.29})--(\ref{2.32}) has allowed 
us to locate the transition temperature between the NP phase ($\eta=0$) and 
the PAIR phase ($\eta \neq 0$)  as a function of the pairing interaction 
strength $g$ and the magnetic field $H_{z}$ (see Fig.\ (1)). The nature of 
the transition line has a complex dependence on both quantities. 

For instance, if $H_{z}$=0, for high $T$ and small $g$ there is no long 
range order $\eta=0$ ($Q$ is always zero) that corresponds to the NP phase. 
On the other hand, for high $T$ and high $g$ the parameter $\eta$ is nonzero. 
In this situation there is pair formation on the sites which corresponds to 
the PAIR phase. The transition between NP and PAIR phases is a second order 
type, which means that the order parameter $\eta$ goes continuously to zero 
as $g$ decreases. That is shown in Fig.\ (2). However, if $g$ is decreased 
bellow a particular value $g=g_{tc}$ (which corresponds to the temperature 
$T=T_{tc}$), the order parameter starts to display a discontinuous behavior 
indicating that the nature of the transition line has changed to a first 
order one. In the absence of the magnetic field the behavior 
of the other parameters are $R$=$R_{z}$ and $Q_{z}=0$. 

If the magnetic field is turned on, the transition NP-PAIR only exists for 
larger values of the pairing strength $g$ and the field tends to destroy the 
PAIR phase (Fig.\ (1)). The tricritical point ($g_{tc}$, $T_{tc}$) is moved 
up and, as a consequence, the first order transition line exists over a larger 
interval of the pairing strength $g$. 

The ($g_{tc}$, $T_{tc}$) point has been confirmed as a tricritical point by 
the expansion of the Gran Canonical Potential in powers of the order 
parameter $\eta$ which defines the symmetry of the pairing phase:
\begin{equation} 
\beta\Omega=\beta\Omega_{0} + A_{1}\eta^{2}+ A_{2}\eta^{4}
\label{3.1}
\end{equation}
where 
\begin{equation} 
A_{1}=\frac{\beta^{2}g^{2}}{2}\int^{+\infty}_{-\infty}Dw\frac{1}
{\int^{+\infty}_{0}ue^{\frac{-u^{2}}{2}}du\int^{+\infty}_{-\infty}
Dv[1+ \cosh(\beta|\vec{h}|]} - 1
\label{3.2}
\end{equation}
\begin{equation} 
A_{2}=\frac{\beta^{4} g^{4}}{64}\int^{+\infty}_{-\infty}Dw
\frac{1-\frac{1}{3}\int^{+\infty}_{0}ue^{\frac{-u^{2}}{2}}du
\int^{+\infty}_{-\infty} Dv[1+ \cosh(\beta|\vec{h}|]}
{\{\int^{+\infty}_{0}ue^{\frac{-u^{2}}{2}}du\int^{+\infty}_{-\infty}
Dv[1+ \cosh(\beta|\vec{h}|]\}^{2}}\;\; .
\label{3.3}
\end{equation}

~

The condition $A_{1}=0=A_{2}$ (together with the equations for $R$, $R_{z}$ 
and $Q_{z}$) gives the precise location of the tricritical point according 
to the known criteria\cite{La}. This location is shown in Fig.\ (\ref{fig1}). 
Tricritical points have been already found in fermionic spin glasses for 
an Ising model with a pairing interaction in the half filling situation\cite{7} 
as well for an Ising model with charge fluctuation\cite{Opp1}. 
 
The behavior of the parameters $R$, $R_{z}$ and $Q_{z}$ also changes in 
the presence of the magnetic field. Fig.\ (\ref{fig2}) shows the results for 
$H_{z}=0.00$ where $Q_{z}=0$ and $R=R_{z}$. Figs. (\ref{fig3}), (\ref{fig4}) 
and (\ref{fig5}) show the results, respectively, for $H_{z}/J/(8)^{1/2}$ = 
0.25, 0.50 and 0.75 where one can see that the parameter $Q_{z}$ is no longer 
null while $R\neq R_{z}$.  


\section{Conclusions} \label{conclu}

In this work we have investigated how the mechanism responsible for the BCS 
pairing formation in real space (PAIR phase) can be affected for a spin 
glass diagonal replica symmetric order parameter through a long ranged 
Heinsenberg model with a pairing interaction in presence of magnetic 
field in $z$ direction. This model could be treated at mean field level 
within the static approximation and the replica symmetry ansatz, always with 
no transversal (to the field) spin glass ordering in order to remain in a 
region where the static approximation and replica symmetry are reliable. So, 
the initial problem has been reduced to one site problem with two different 
effective fields applied on, which are $\eta$ ($\eta^{*}$) and $\vec{h}$. 
The former is a long range internal field related to a symmetry breaking 
which produces a pairing long range order and appears in the problem combined 
with the chemical potential to give an effective chemical potential. For the 
half filling case, just the pairing field survives. The later is a random 
gaussian field related to $R$, $R_{z}$ and $Q_{z}$ as defined in Sec.\ 
\ref{gen_form}. As we can see in  Eqs.\ (\ref{2.29})--(\ref{2.32}), the 
effective fields have to be solved coupled. That would be the mechanism 
through which the magnetic part of the problem always affects the PAIR phase, 
even above of the spin glass transition. 

\begin{figure}
\vspace*{-1.5cm}
\hspace*{-1.5cm}
\begin{minipage}[c]{12cm}
\includegraphics[angle=0,scale=0.85]{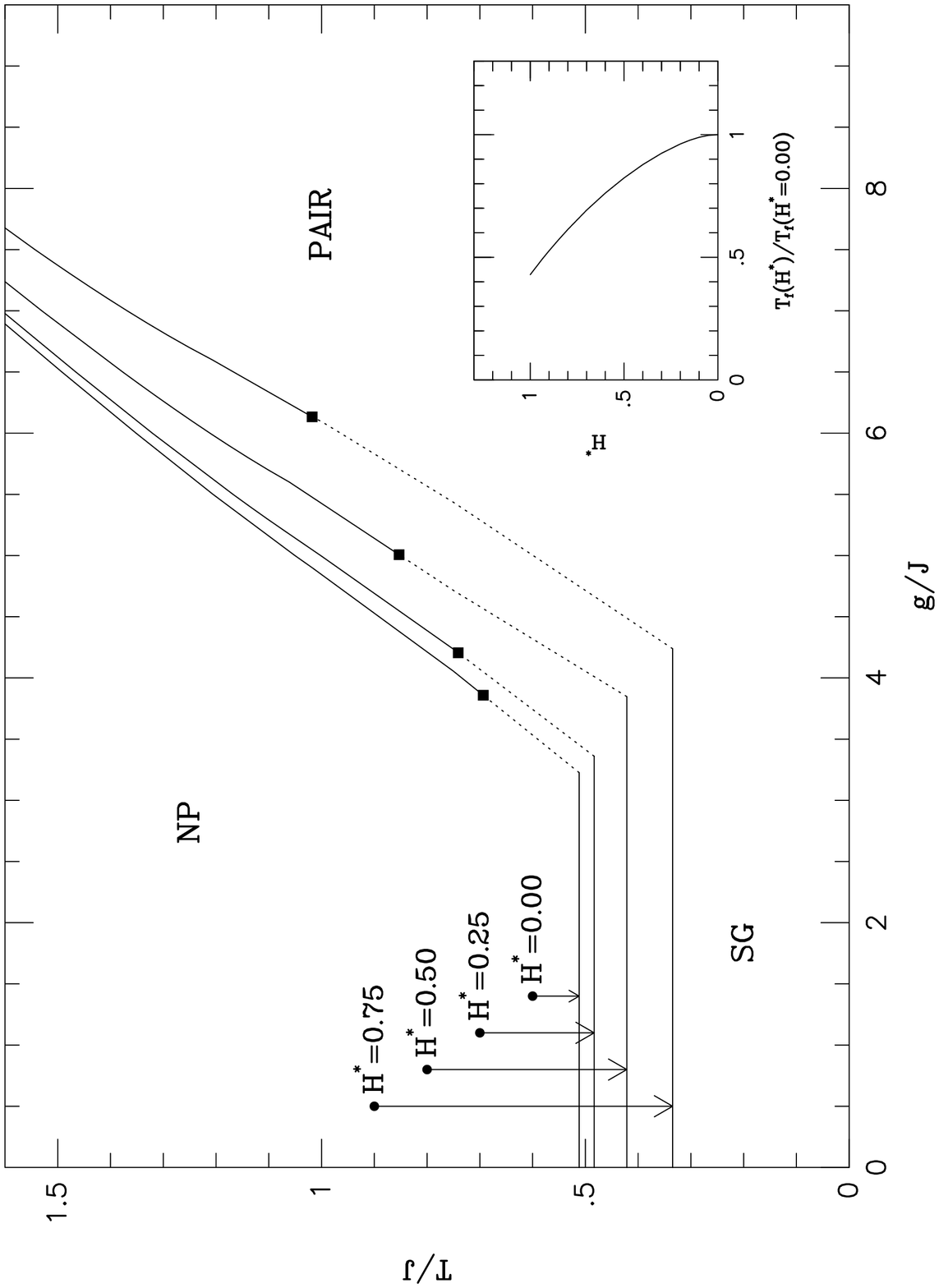}
\end{minipage}
\hspace*{4.8cm}
\begin{minipage}[c]{1cm}
\vspace*{11mm}
\rotcaption{Phase diagram as a function of temperature and pairing coupling 
$g/J$ for several values of $H^{*}$ where $H^{*}=H_{z}/J/(8)^{1/2}$. 
Solid lines indicate second order transitions while dotted lines indicate 
first order transitions. Tricritical points are shown as filled squares. 
In the lower right corner it is shown the relation between $H^{*}$ and the 
NP--SG transition temperature $T_{f}$ relative to $T_{f}(H^{*}=0.00)$. 
The transition line between SG--PAIR is not attainable from the present 
approach.}
\label{fig1}
\end{minipage}
\vspace*{-24.5cm}
\begin{center} \huge\bf Figure \ref{fig1} \end{center}
\end{figure}

\begin{figure}
\vspace*{-2cm}
\hspace*{-1cm}
\includegraphics[width=18cm]{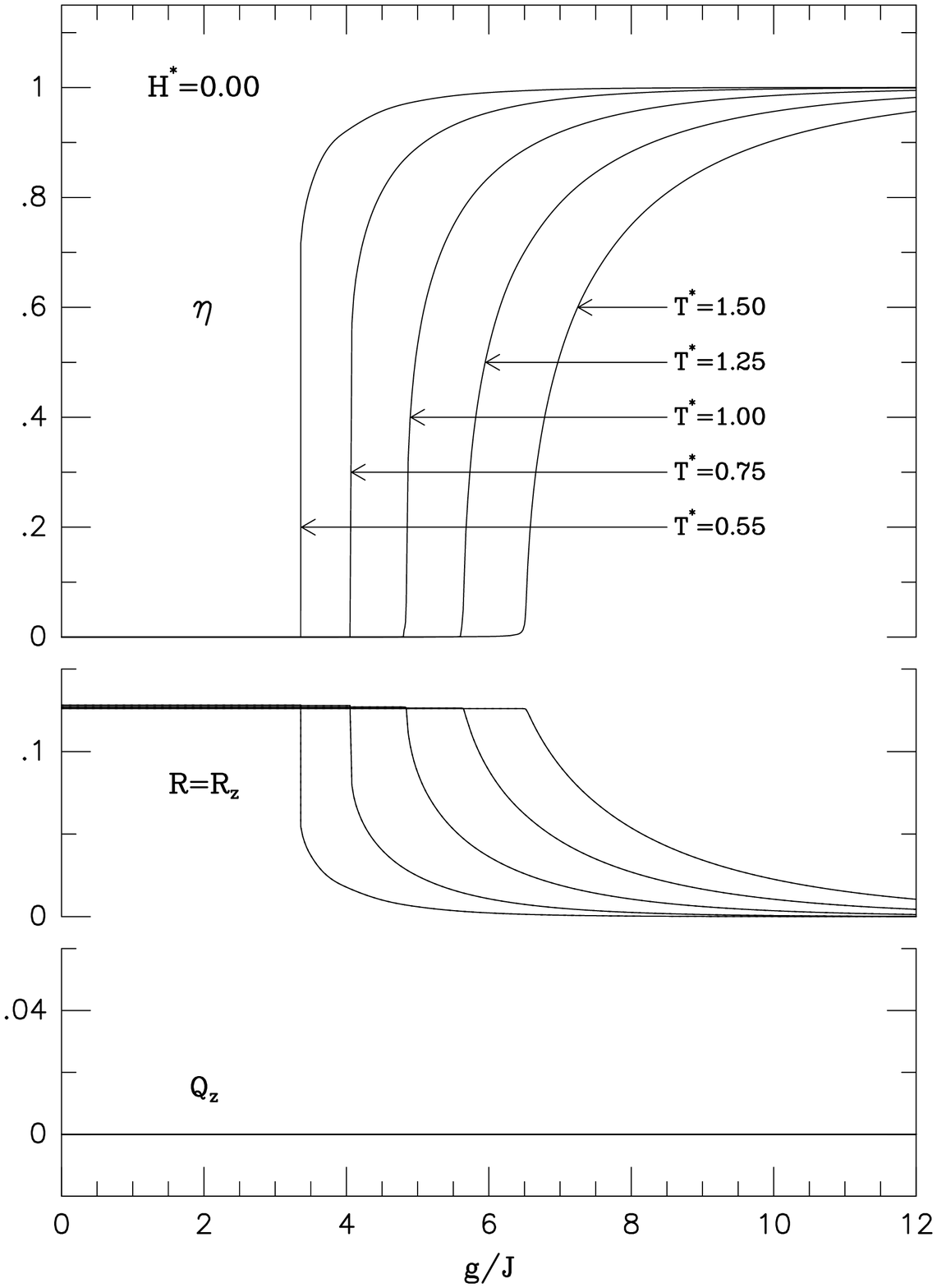}
\vspace*{-13mm}
\caption{Dependence of the order parameters $\eta$, $R$, $R_z$ and $Q_z$ for 
$H^{*}=0.00$ as a function of $g/J$ for several values of temperature where
$T^{*}=T/J$. For high values of $g/J$ the transition is clearly a continuous 
one as opposed to a discontinuous one for lower values of $g/J$. 
For $H^{*}=0.00$, $Q_z$ is always null.} 
\label{fig2}
\vspace*{-24.5cm}
\begin{center} \huge\bf Figure \ref{fig2} \end{center}
\end{figure}

\begin{figure} 
\vspace*{-2cm}
\hspace*{-1cm}
\includegraphics[width=18cm]{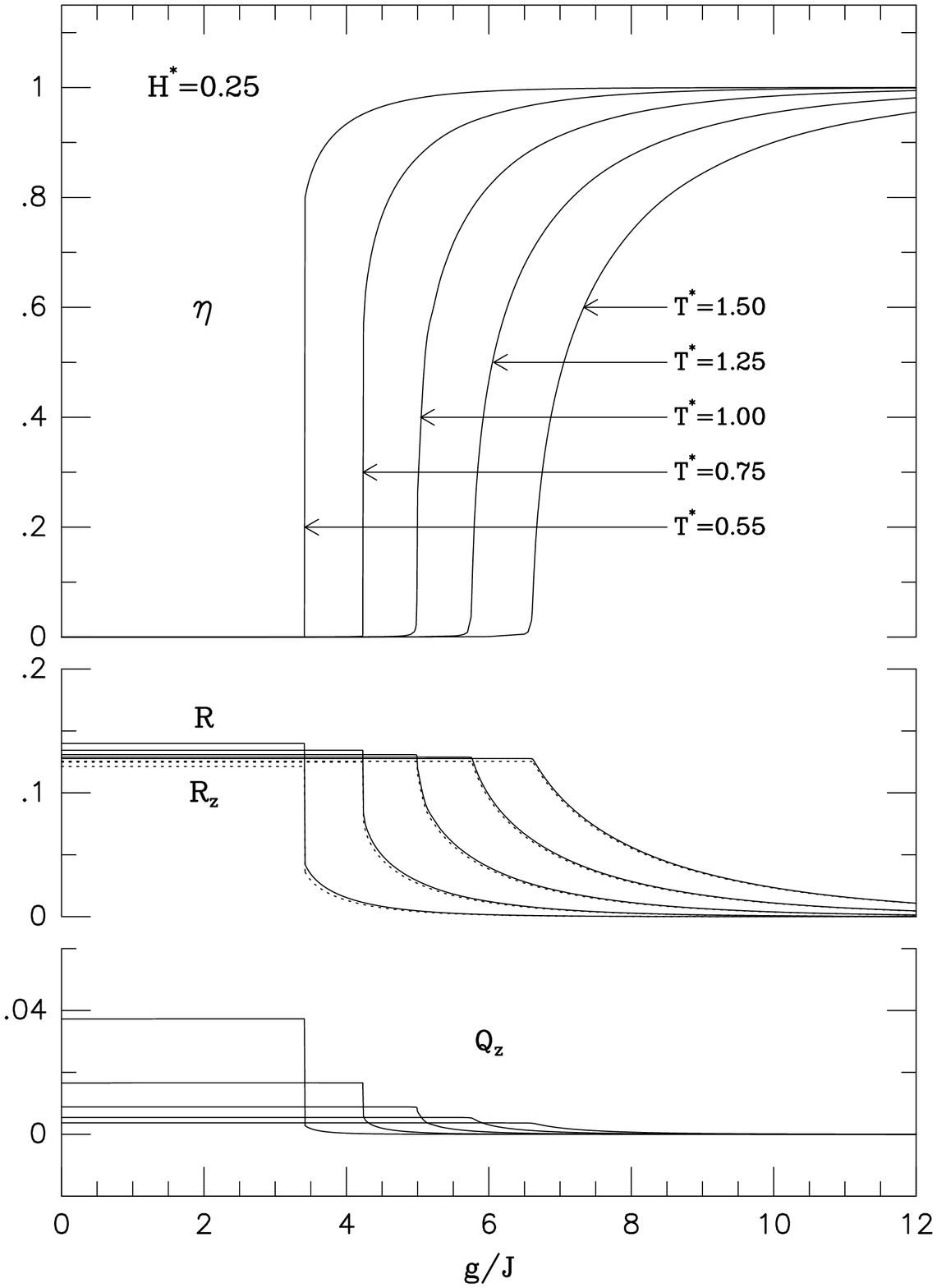}
\vspace*{-13mm}
\caption{Same as Fig.\ \ref{fig2} for $H^{*}=0.25$.} 
\label{fig3}
\vspace*{-23.8cm}
\begin{center} \huge\bf Figure \ref{fig3} \end{center}
\end{figure}

\begin{figure} 
\vspace*{-2cm}
\hspace*{-1cm}
\includegraphics[width=18cm]{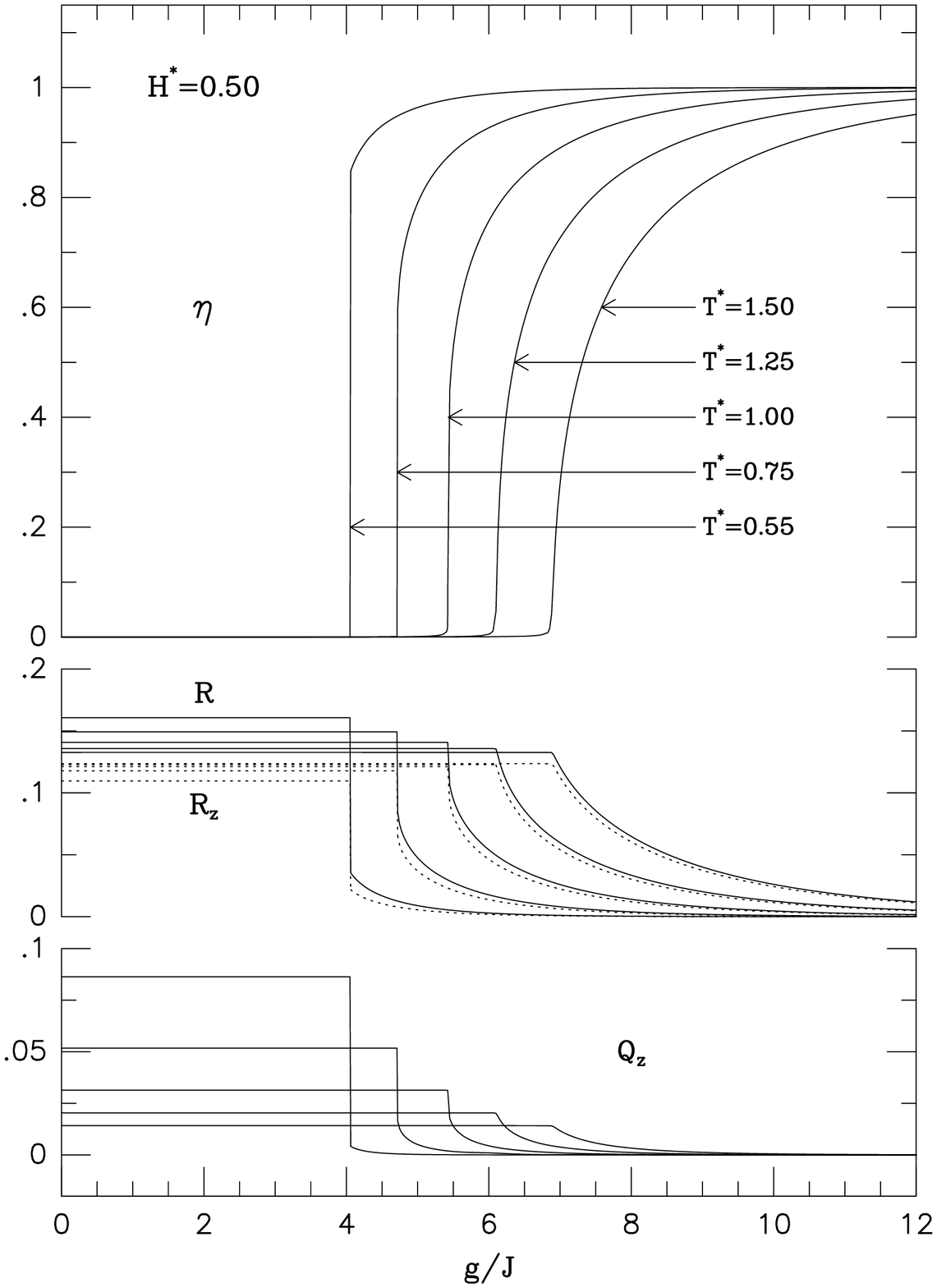}
\vspace*{-13mm}
\caption{Same as Fig.\ \ref{fig2} for $H^{*}=0.50$.} 
\label{fig4}
\vspace*{-23.8cm}
\begin{center} \huge\bf Figure \ref{fig4} \end{center}
\end{figure}

\begin{figure}
\vspace*{-2cm}
\hspace*{-1cm}
\includegraphics[width=18cm]{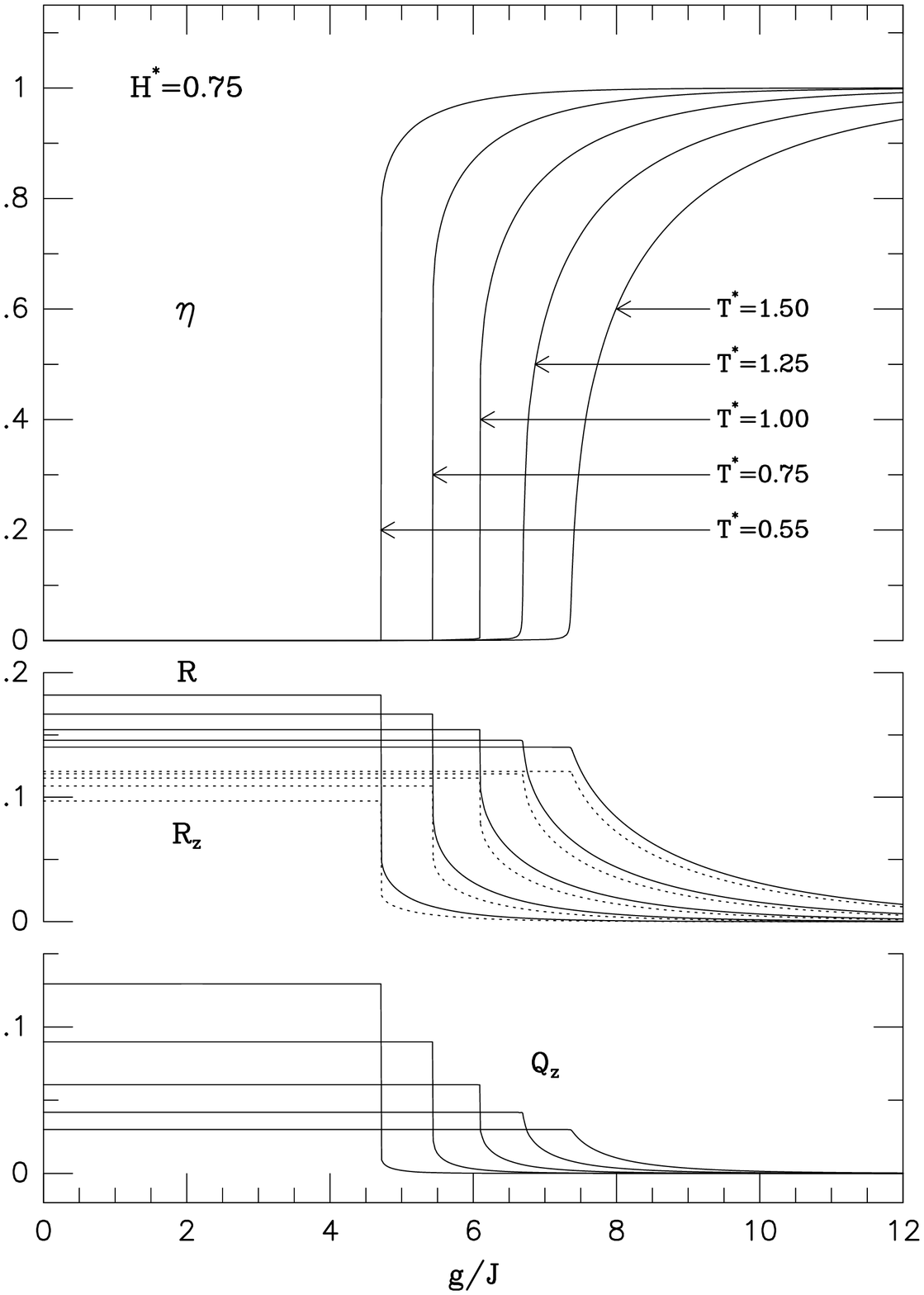}
\vspace*{-13mm}
\caption{Same as Fig.\ \ref{fig2} for $H^{*}=0.75$.} 
\label{fig5}
\vspace*{-23.8cm}
\begin{center} \huge\bf Figure \ref{fig5} \end{center}
\end{figure}

By solving numerically the Eqs.\ (\ref{2.29})--(\ref{2.32}), it has been 
possible to construct the phase diagram for several values of $H_{z}$ 
(Fig.\ (\ref{fig1})). For high temperatures, it has been found a crossover 
from a continuous pair breaking symmetry to a sharp one where the PAIR order 
parameter $\eta$ displays a discontinuous behavior as shown in 
Figs.\ (\ref{fig2}--\ref{fig5}). The presence of the field $H_{z}$ creates 
an anisotropy which can be seen clearly in the behavior of the $R$, $R_{z}$ 
and $Q_{z}$, shown in Figs.\ (\ref{fig3})--(\ref{fig5}). Therefore, the 
region in the phase diagram where there is a first order transition becomes 
larger. This behavior could be verified by positioning the tricritical point 
using the Eqs.\ (\ref{3.2}) and (\ref{3.3}).

We remark that the our fermionic spin representation constitutes an important
difference from Ref.\ \onlinecite{9}. The consequences of different 
representations for the spins are clear if one compares our equations 
for the order parameters with those obtained in Ref.\ \onlinecite{9}. We also 
point out that the transition line between SG--PAIR is not attainable from the 
theory used in this work. 

Lastly, we have worked within the scope of the static approximation. The 
dynamic is absent in this formulation and certainly it is responsible for 
important effects at very low temperatures. Nevertheless, the static 
approximation is an upper bound of the theory from where the dynamic should 
be properly included. That would be subject for future work.

         
\acknowledgements We are grateful to Prof. Alba Theumann for relevant 
comments. The numerical calculations were partially performed at LANA, 
Departamento de Matem\'{a}tica, CCNE, UFSM. This work was partially supported 
by FAPERGS (Funda\c{c}\~{a}o de Amparo \`{a} Pesquisa do Rio Grande do Sul).



\newpage

\newpage

\newpage

\newpage

\newpage

\begin{references}
%
\bibitem{1} 
H.\ Spille, M.\ Winkelmann, U.\ Ahlheim, C.\ D.\ Bredl,
F.\ Steglich, P.\ Haen, J.\ M.\ Mignot, J.\ L.\ Tholence and R.\ Tournier,
J.\ Magn.\ Magn., {\bf 76/77 }, 539 (1988).
%
\bibitem{2}
S.\ Bath, H.\ R.\ Ott, F.\ N.\ Gygax, B.\ Hitti, E.\ Lippelt,
A.\ Schenk, J.\ Magn.\ Magn., {\bf 76/77 }, 455 (1988).
%
\bibitem{3}
F.\ C.\ Chou,  N.\ R.\ Belk, M.\ A.\ Kastner, R.\ J.\ Biergenau and 
A.\ Aharony, Phys.\ Rev.\ Lett., {\bf 75}, 2204 (1995).
%
\bibitem{4}
D.\ J.\ Scalapino, Phys.\ Rep., {\bf 250}, 329 (1995). 
%
\bibitem{5}
D.\ Davidov, K.\ Baberschke, J.\ A.\ Mydosh and G.\ J.\ Nieuwenhuys, J.\ 
Phys.\ F.: Metal.\ Phys., {\bf 7} L47 (1977).
%
\bibitem{6}
M.\ J.\ Nass, K.\ Levin and G.\ S.\ Grest, Phys.\ Rev. B {\bf 23} 1111 (1981).
%
\bibitem{She}
D.\ Sherrington and M.\ Simkin, J.\ Phys.\ A: Math.\ Gen., {\bf 26}, 
L1201 (1993).
%
\bibitem{Jose}
T.\ K.\ Kop\'{e}c and J.\ V.\ Jos\'{e}, Phys.\ Rev.\ B, {\bf 52}, 16140 
(1995-I).
%
\bibitem{Opp}
H.\ Feldmann and R.\ Oppermann, cond-mat/9809001.
%
\bibitem{Sach} 
a) S.\ Sachdev and J.\ Ye, Phys.\ Rev.\ Lett., {\bf 70} 3339 (1993). \\
b) J.\ Ye, S.\ Sachdev and N.\ Read, Phys.\ Rev.\ Lett., {\bf 70} 4011 (1993). 
%
\bibitem{Ge}
Anirvan M.\ Sengupta, Antoine Georges, Phys.\ Rev.\ B, {\bf 54}, 10295 
(1995-II).
%
\bibitem{7}
S.\ G.\ Magalh\~{a}es and Alba Theumann, Europ.\ Phys.\ 
J.\ B, {\bf 9}, 5 (1999).
%
\bibitem{8}
A.\ J.\ Bray and M.\ A.\ Moore, J.\ Phys.\ C, {\bf 13}, 419 (1980).
%
\bibitem{Alb}
Alba Theumann, Phys.\ Rev.\ B, {\bf 33}, 559 (1986). 
%
\bibitem{9}
Yadin Y.\ Goldshmidt and Pik-Yin Lai, Phys.\ Rev.\ B, {\bf 43}, 11434 (1991).
%
\bibitem{10}
K.\ A.\ Penson and M.\ Kolb, Phys.\ Rev.\ B {\bf 33}, 1663 (1986).
%
\bibitem{11}
M.\ Kolb and K.\ A.\ Penson, J.\ Stat.\ Phys., {\bf 44}, 129 (1986).
%
\bibitem{12}
a) D.\ Sherrington and S.\ Kirkpatrick, Phys.\ Rev.\ Lett., {\bf 35}, 1972
  (1979).\\ 
b) S.\ Kirkpatrick and D.\ Sherrington, Phys.\ Rev.\ B, {\bf 17}, 4384 (1978).
%
\bibitem{13}
a) John W.\ Negele and Henry Orland, {\it  Quantum Many Particles}, 
   1987 (Addison--Wesley Publishing Company). \\
b) K.\ B.\ Efetov, Adv.\ in Physics, {\bf 32}, 53 (1983). \\
c) D.\ Sherrington, in {\it Path Integrals, NATO Advanced Institute Series,    
   Editor: G.\ J.\ Papadopoulos and J.\ T.\ Devreese}, \\ \phantom{c) }Plenum 
   Press, New York, 1978.
%
\bibitem{MUS}
B.\ M\"{u}hlschlegel, J.\ of Math.\ Phys., {\bf 3}, 522 (1962).
%
\bibitem{ALB2}
Alba Theumann, Phys.\ Rev.\ B, {\bf 11}, 4382 (1972).
%
\bibitem{La}
L.\ D.\ Landau and E.\ M.\ Lifshitz, {\it Statistical Physics} $3^{rd}$ 
Edition, Pergamon Press, 1980.
%
\bibitem{Opp1}
B.\ Rosenow and R.\ Oppermann, Phys.\ Rev.\ Lett., {\bf 77}, 1608 (1996).
%
\end{references}
\end{document}